\documentclass[a4paper,fleqn]{cas-dc}
\usepackage[round]{natbib}

\usepackage{eurosym,makecell,comment}
\usepackage{multirow}
\usepackage{threeparttable}
\usepackage{framed}
\usepackage{hyperref}
\usepackage{url}
\usepackage{verbatim}
\usepackage{graphicx}
\usepackage{subcaption}
\usepackage{float}
\usepackage{xcolor}
\usepackage{microtype}
\usepackage{colortbl}
\usepackage{amsmath}
\usepackage{mathtools}

\usepackage{array} 

\usepackage{etoolbox}
\usepackage{soul} 

\newtoggle{finalPaper}

\setstcolor{red}

\toggletrue{finalPaper} 

\iftoggle{finalPaper} {
	\newcommand{\addtxt}[1]{#1}
	\newcommand{\change}[2]{#2}
	\newcommand{\rmvtxt}[1]{}}
{
	\newcommand{\addtxt}[1]{\textcolor{red}{#1}}
	\newcommand{\change}[2]{\st{#1}\textcolor{red}{#2}}
	\newcommand{\rmvtxt}[1]{\st{#1}}}

\begin{document}
\let\WriteBookmarks\relax
\def\floatpagepagefraction{1}
\def\textpagefraction{.001}
\shorttitle{AuthCODE: A Privacy-preserving and Multi-device Continuous Authentication Architecture based on Machine and Deep Learning}
\shortauthors{S\'anchez et~al.}

\title[mode = title]{AuthCODE: A Privacy-preserving and Multi-device Continuous Authentication Architecture based on Machine and Deep Learning}

\author[1]{Pedro Miguel S\'anchez S\'anchez}[orcid=0000-0002-6444-2102]
\cormark[1]

\author[2]{Lorenzo Fern\'andez Maim\'o}[orcid=0000-0003-2027-4239]

\author[3]{Alberto Huertas Celdr\'an}[orcid=0000-0001-7125-1710]

\author[1]{Gregorio Mart\'inez P\'erez}[orcid=0000-0001-5532-6604]

\address[1]{Department of Information and Communications Engineering, University of Murcia, Murcia 30100, Spain}

\address[2]{Department of Computer Engineering, University of Murcia, Murcia 30100, Spain}

\address[3]{Communication Systems Group (CSG), Department of Informatics (IfI), University of Zurich UZH, 8050 Zürich, Switzerland}

\cortext[cor1]{Corresponding author.
Email address: pedromiguel.sanchez@um.es (P.M.S. S\'anchez)}


\maketitle      

\begin{abstract}
The authentication field is evolving towards mechanisms able to keep users continuously authenticated without the necessity of remembering or possessing authentication credentials. While relevant limitations of continuous authentication systems -high false positives rates (FPR) and difficulty to detect behaviour changes- have been demonstrated in realistic single-device scenarios, the Internet of Things and next generation of mobile networks (5G) are enabling novel multi-device scenarios, such as Smart Offices, that can help to reduce or address the previous challenges. The paper at hand presents an AI-based, privacy-preserving and multi-device continuous authentication architecture called AuthCODE. AuthCODE seeks to improve single-device solutions limitations by considering additional behavioural data coming from heterogeneous devices. AuthCODE proposes a novel set of features that combine the interactions of users with different devices. The features relevance has been demonstrated in a realistic Smart Office scenario with several users that interact with their mobile devices and personal computers. In this context, a set of single- and multi-device datasets have been generated and published to compare the performance of our multi-device solution against single-device approaches. A pool of experiments with machine and deep learning classifiers measured the impact of time in authentication accuracy and improved the results of single-device approaches by considering multi-device behaviour profiles. Specifically, the multi-device approach using XGBoost with 1-minute window of aggregated features, achieved a 69.33\%, 59,65\% and 89,35\% improvement in the FPR when compared to the single-device approach for computer, mobile applications and mobile sensors respectively.
Finally, temporal information classified by a Long-Short Term Memory Network, allowed the identification of additional complex behaviour patterns.

\end{abstract}

\section{Introduction}

Continuous authentication systems pretend to improve the limitations of traditional mechanisms, which authenticate users from time to time according to credentials such as passwords, codes, or tokens \citep{almalki2019continuous}. In this context, continuous authentication mechanisms increase the level of security, keeping users authenticated permanently, and enhance the users' quality of experience (QoE), being non-intrusive and minimizing the usage of credentials during the authentication processes~\citep{ContAuth}.

Existing continuous authentication mechanisms model the user's behaviour when he/she uses a particular device for a given time. \addtxt{We understand by "behaviour" the set of actions that a given user performs somewhat consciously with one or more devices. As an example, for PC devices the user's behaviour could be determined by the mouse movements, keystrokes, or applications opened and closed. Regarding mobile devices, the behaviour might be related to screen interactions, device movements, or patterns to use applications. From a multi-device perspective, the user's behaviour could be characterized by the device used at each particular moment of the day, the duration that each device is used, or the type of application used in each device. }As a result \addtxt{of user monitoring}, a behaviour profile is created and stored in a dataset, being the precise selection of dimensions, data and features critical to create accurate behaviour profiles. The next step is to use this profile to train Machine Learning (ML) or Deep Learning (DL) models, which can be classifiers or anomaly detectors depending on the nature of the data and the approached scenario. Finally, the continuous authentication mechanism uses these models to evaluate the similarity between the current device usage profile and the learned user's profile, providing either a user's identification or an anomaly score which can be used to decide if the user is authenticated or not.

Different solutions have implemented the previous steps for single-device scenarios such as personal computers and smartphones \citep{Sensors}. However, they present some important limitations such as the high rate of false positives, which happens when users change some aspect of their behaviour and the system does not recognize them\change{, or the impossibility of detecting some impersonation attacks}{. The lack of privacy considerations and the impossibility of detecting some impersonation attacks are two additional limitations of single-device solutions}. In this context, the evolution of information and communications technology such as 5G networks \citep{5G} or the Internet of Things (IoT) \citep{IoT} is influencing the applicability of continuous authentication in multi-device scenarios to reduce the limitations of single-device scenarios by including additional information coming from the user's activity on diverse devices used during a time window. On the one hand, multi-device continuous authentication mechanisms could consider data from different devices to decide if the user changed his/her behaviour, which could reduce the false positive rate. On the other hand, multi-device scenarios such as Smart Offices \citep{Smart_Office} could also get a benefit from non-invasive and robust multi-device continuous authentication, using them to control the access to sensitive data managed by heterogeneous devices such as IoT devices, tablets, smartphones, or computers. \addtxt{As an example, if an attacker studies the common activities and routines of a user protected by a single-device solution, this attacker can learn his behavior and then mimic him to some extent, performing malicious activities.} \addtxt{However, having information from several devices improves the robustness and flexibility of the authentication system since it would be necessary to have access to several devices and mimic the user's behaviour in each of them.}

Despite the benefits of existing continuous authentication solutions, their design, implementation and integration in new multi-device scenarios are open and challenging issues. In this sense, we emphasize the following challenges: 1) what features and data are relevant to create collaborative rich users' behaviour profiles in multi-device scenarios; 2) whether multi-device continuous authentication mechanisms can improve the performance of single-device mechanisms in terms of false positive and negatives; \rmvtxt{and }3) how the number or interactions and time affect the accuracy of continuous authentication mechanisms\addtxt{; and 4) how behavioural data is collected and processed in order to guarantee users' privacy}.



The main contributions of this paper are the following ones:
\hyphenation{pa-ra-digms}
\begin{itemize}
    \item A multi-device continuous authentication architecture, called AuthCODE, that guarantees the privacy of users' sensitive data while improving the authentication performance of single-device solutions. AuthCODE considers a hybrid approach that combines the Mobile Edge Computing (MEC) and Cloud Computing paradigms. Privacy-preserving features are calculated in the MEC and sent to the cloud, where ML/DL models are trained and evaluated to keep the authentication performance. 
    
    \item A set of privacy-preserving and multi-device features able to model precisely the users' behaviour when they interact, in a collaborative way, with heterogeneous devices during different time windows. \addtxt{In this way, the user's activities can be accurately characterized, but if the resultant dataset falls into the hands of an attacker, it would be harmless as the user's monitored activities do not contain sensitive information such as passwords or other typed text.}
    
    \item Five datasets, available in \citep{datasets}, which contain single- and multi-device privacy-preserving features modelling the users' behaviour. The datasets have been created by modelling a realistic Smart Office scenario where five users interact with their computers and mobile devices. Regarding computer interaction, the data aggregate information about keys pressed, mouse actions, as well as application and resource usage statistics. With regard to mobile devices, the data consist of sensor values and application usage statistics.
    
    \item A pool of experiments with ML and DL classifiers performed over the previous five datasets that demonstrated how the multi-device behaviour profile of AuthCODE improves the false positive rates (FPR) and f1-score metrics of our single-device approach. Specifically, the f1-score and FPR average reached for multi-device profiles of 1-minute windows were 99.33\% and 0.23\%, respectively, while the same metrics for single-device authentication were 97.39\% and 0.72\% in the case of personal computers, 96.70\% and 0.57\% for mobile app statistics, and 90.36\% and 2.16\% for mobile sensors. Therefore, the improvement achieved in FPR was 69.33\%, 59,65\% and 89,35\% for computers, mobile app statistics and mobile sensors respectively. Finally, to take advantage of the complex patterns present in multi-device profiles, a set of Long-Sort Term Memory (LSTM) neural network configurations obtained an average f1-score of 89\% and FPR of 2.02\% for a 5-minute sliding window and +99\% and 0.37\%, respectively, using +60-minute sliding window.
    
\end{itemize}

The remainder of the paper is organized in the following way. Section~\ref{related} discusses the related work focused on continuous authentication for multi-device scenarios as well as single-device such as personal computers and mobile devices. A motivating use case is detailed in Section~\ref{sec:motivating}. The architectural design of AuthCODE is explained in Section~\ref{sec:architecture}. The implementation details of the proposed solution as well as a realistic Smart Office scenario are explained in Section~\ref{sec:deployment}. Section~\ref{sec:experiments} measured the performance of AuthCODE in the Smart Office environment. Finally, Section~\ref{conclusions} shows the conclusions and future work.

\section{Related Work} 
\label{related}

This section reviews the main continuous authentication solutions found in the literature. A wide variety of continuous authentication proposals focused on single- and multi-device scenarios are identified and analysed, extracting common ideas and potential improvements. 

\begin{table*}

	\caption{Continuous authentication solutions comparison.}
	\centering
    \begin{tabular}{m{2.3cm}cm{2.9cm}m{2.5cm}m{6.3cm}}
	    \hline
	    \makecell[c]{\textbf{Proposal}} & \makecell[c]{\textbf{Device}\\\textbf{Type}} & \makecell[c]{\textbf{Dimensions}} & \makecell[c]{\textbf{Algorithms}} & \textbf{\makecell[c]{Results /\\ Conclusions}} \\
		\hline
		\citep{Sensors} & \makecell[c]{Mobile} & \makecell[c]{Sensors and \\Application Usage\\ Statistics} & \makecell[c]{Isolation\\Forest} & \makecell[c]{Adaptability, low resource consumption, 92\%\\ Recall and 77\% Precision, and resilience to \\adversarial attacks} \\
		\hline
		\citep{BoAuth2013} & \makecell[c]{Mobile} & \makecell[c]{Sensors and \\touchscreen events} & \makecell[c]{One Class - SVM\\ SVM} & \makecell[c]{72.36\% Accuracy and 24.99\% FAR}\\
		\hline			
		 \citep{PatelAuth2016}& \makecell[c]{Mobile} & \makecell[c]{Survey} & \makecell[c]{-} & \makecell[c]{Review about current state and challenges}\\
		 \hline
		 \citep{LiAuth2018}& \makecell[c]{Mobile} & \makecell[c]{Sensors qith\\data augmentation} & \makecell[c]{One-Class SVM} & \makecell[c]{7.65\% FAR, 9.01\% FRR and 8.33\% EER.}\\
        \hline
		\citep{FridmanAuth2017} & \makecell[c]{Mobile} & \makecell[c]{Text, location, \\application usage\\ and websites} & \makecell[c]{SVM} & \makecell[c]{95\% Accuracy and 5\% EER}\\
        \hline
		\citep{CentenoAuth2017} & \makecell[c]{Mobile} & \makecell[c]{Sensors} & \makecell[c]{Autoencoder} & \makecell[c]{97.8\% Accuracy and 2.2\% EER} \\
        \hline
		\citep{EhatishamAuth2017} & \makecell[c]{Mobile}  & \makecell[c]{Sensors} & \makecell[c]{Bayes Net and\\ Euclidean distance} & \makecell[c]{87.34–90.78\% Accuracy}\\
        \hline
		\citep{DeutschmannBehavioral2013} & \makecell[c]{Desktop}  & \makecell[c]{Mouse, Keyboard\\ and Used applications} & \makecell[c]{Bayes Net} & \makecell[c]{18 seconds to detect an imposter using \\ 15-50 keystrokes and 2.4 mins using \\66 mouse interactions}\\
		\hline
		 \citep{FridCon2015} & \makecell[c]{Desktop} & \makecell[c]{Keyboard and \\mouse events\\ linked to application.} & \makecell[c]{Naive Bayes\\ and SVM} & \makecell[c]{0.4\% FAR and 1\% FRR after 30~s \\ 0.1\% FAR and 0.2\% FRR after 5 minutes}\\ 
		\hline
		 \citep{AljohaniConAIS2018} & \makecell[c]{Desktop} & \makecell[c]{Keyboard and mouse} & \makecell[c]{Artificial Immune\\System (AIS)} & \makecell[c]{97.05\% average Accuracy (96.6\% to 97.74\%)}\\
		\hline
		\citep{MondalAuth2017} & \makecell[c]{Desktop} & \makecell[c]{Keyboard and mouse} & \makecell[c]{Decision Tree, \\N. Network, SVM} & \makecell[c]{Many experiments and test performed. \\0.04-1\% EER}\\
		\hline
		\citep{lu2020continuous} & \makecell[c]{Desktop} & \makecell[c]{Keyboard} & \makecell[c]{CNN, RNN} & \makecell[c]{2.07\% and 6.61\% FAR, 3.26\% and\\ 5.31\% FRR, and 2.67\% and  5.97\% EER\\, for CNN and RNN, respectively} \\
		\hline
		\citep{SanchezAuthIoT2019} & \makecell[c]{IoT} &\makecell[c]{Mouse and keyboard \\(PC), Application \\usage (Mobile)} & \makecell[c]{Random Forest} &  \makecell[c]{97.43\% precision, 96.20\% recall, 96.76\% \\F1-Score (Mobile), 96.32 \% precision, 90.00\% \\recall, 92.70\% F1-Score (PC)} \\
		\hline
		 \citep{AshibaniAuthIoT2019} & \makecell[c]{Smart\\Home} & \makecell[c]{User, Device, \\Network and \\Environment context} & \makecell[c]{Heuristic analysis \\and pattern\\ matching} & \makecell[c]{$<$100 ms authentication time and verified \\improvement over using only credentials} \\
		\hline
	    \citep{NespoliAuthIoT2018}& \makecell[c]{IoT} & \makecell[c]{Location, Person and \\IoT Devices \\Ontologies} & \makecell[c]{Ontologies and\\ semantic rules} & \makecell[c]{Execution time less than 4s and more \\than 78\% mean confidence level}\\
		\hline
		AuthCode (Ours) & \makecell[c]{IoT} & \makecell[c]{Modular architecture \\(PC and mobile)} & \makecell[c]{MLP, XGBoost, \\RF and LSTM} & \makecell[c]{Precision: 99.32\%, Recall: 99.33\%, F1-Score:\\ 99.33\% (more results in experiments section)} \\
		\hline
	\end{tabular}
	\label{tab:discussion}
\end{table*}

\subsection{Continuous authentication in single-device scenarios}

In the literature, we can find the next two families of single-device scenarios: mobile devices and personal computers.

In the field of continuous authentication for mobile devices, Jorquera Valero et al.~\citep{Sensors} proposed a continuous and adaptive authentication system based on monitoring the application usage statistics and device sensors (gyroscope and accelerometer). The authors used ML-based anomaly detection techniques, concretely Isolation Forest, to identify anomalies in the users' behavioural data. Each user dataset is dynamically updated using his/her current behaviour in order to achieve system temporal adaptability. The solution obtained 92\% recall and 77\% precision when authenticating different users. In \citep{BoAuth2013}, Bo et al. identified users thanks to biometrics and typing patterns. The proposed system considered rotation, vibration, and pressure of smartphone touchscreens and sensors. One-class SVM were used over the user behaviour to classify his profile, achieving 72.36\% accuracy and 24.99\% FAR (False Acceptance Rate). In \citep{PatelAuth2016}, Patel et al. performed a review about the definition of continuous authentication and the proposals in the field. Authors focused on dimensions and AI techniques that are applied commonly in continuous authentication. They mentioned facial recognition, gestures, application usage and location, and concluded that merging data from different dimensions results in better accuracy and lower error rates. Li et al. \citep{LiAuth2018} proposed the application of data augmentation techniques on behavioural information gathered from device sensors (gyroscope and accelerometer) to improve the continuous authentication results. Then, a one-class SVM model was trained and used to evaluate the user, obtaining 7.65\% FAR (False Acceptance Rate), 9.01\% FRR (False Rejection Rate) and 8.33\% EER (Equal Error Rate). Other of the main works in this field is \citep{FridmanAuth2017}, proposed by Fridman et al. This solution uses the typing stylometry, device location, application usage, and accessed websites. Authors achieved 0.4\% FAR and 1\% FRR after 30 seconds. Centeno et al. \citep{CentenoAuth2017} applied Autoencoders in their continuous authentication solution. This solution utilises sensor data to extract device holding patterns. Authors obtain better performance and resource consumption by using a cloud platform to perform the authentication process, achieving 2.2\% EER. In \citep{EhatishamAuth2017}, authors utilised motion sensors in their continuous authentication system. Using these data, several device spatial positions were determined. After performing diverse tests, authors found that Bayes Net was the most appropriate algorithm for position recognition. Then, Euclidean distance was used to evaluate a instance compared to its recognised position, obtaining 87.43\%-90.78\% accuracy (depending on the evaluated position).

In terms of continuous authentication focused on computers and desktop devices, Deutschmann et al. \citep{DeutschmannBehavioral2013} selected keyboard, mouse and usage of applications as representative sources to identify users. Then, classification algorithms were used to sort and filter the gathered information in different categories. The results showed that intruders were detected in 2.4 minutes using the mouse, in 18 seconds using the keyboard, and 1.5 seconds using applications. Lex Fridman et al.~\citep{FridCon2015} utilised keyboard and mouse interactions to identify users. This work proposes to link the keyboard interactions and the application running on the foreground in order to obtain additional information that enables the authentication process. The system used Naive Bayes as classifier for mouse and keyboard events, and SVM for user typing patterns. Using short user interaction periods (30 secs), the system obtained a False Acceptance Rate (FAR) of 0.4\% and a False Rejection Rate (FRR) of 1\%. This metrics decreased to 0.1\% and 0.2\% respectively after a 5 minute evaluation. The system proposed by Aljohani et al. in ~\citep{AljohaniConAIS2018} used an Artificial Immune System (AIS) to perform the continuous authentication task. They used the AIS Negative Selection (NS) algorithm on keyboard and mouse data. After a initialisation period, the system used AIS NS to evaluate keyboard and mouse interaction sets. This solution was evaluated in a group of 24 people, achieving 97.05\% average accuracy (from 97.74\% to 96.6\% in all users). Mondal and Bours~\citep{MondalAuth2017} used keystroke and mouse movement dynamics to build a trust model. Then, this trust model is utilised to perform user continuous authentication. The behaviour of 53 different users was monitored in uncontrolled conditions and then the system was tested using the obtained information. Different Machine Learning classification algorithms were utilised to build a trust model. This trust model utilises a threshold over the dynamic user authentication score. Last, Lu et al. \citep{lu2020continuous} applied Convolutional Neural Networks (CNN) and Recurrent Neural Networks (RNN) to keyboard interactions in computers, achieving 2.07\% and 6.61\% FAR, 3.26\% and 5.31\% FRR, and 2.67\% and 5.97\% EER with CNN and RNN, respectively.

\subsection{Continuous authentication in multi-device scenarios}

This section analyses continuous authentication research works applied to multi-device scenarios. These solutions are the closest to the scope of this work, but the number of existing works is not very high due to the field novelty.

S\'anchez et al. \citep{SanchezAuthIoT2019} designed a continuous authentication architecture oriented to smart offices. Authors deployed their architecture as a proof of concept on mobile phones and computer devices. Then, the architecture was tested separately on the different devices using Random Forest (RF). In mobile, it achieved 97.43\% average precision, 96.20\% average recall and 96.76\% average f1-score. In computer, it achieved 96.32\% average precision, 90.00\% average recall and 92.70\% average f1-score. However, the authors did not perform any experiment combining the users' behaviour in different devices. Ashibani et al. \citep{AshibaniAuthIoT2019} proposed a continuous authentication framework designed for Smart Homes. The framework employed contextual information to authenticate users. It authenticates by considering the user's context, the device context, the network context, and the environmental context, obtained from the smart home IoT devices. Nevertheless, unlike our work, this proposal does not consider user behaviour involving several devices. Another multi-device solution was proposed by Nespoli et al. in~\citep{NespoliAuthIoT2018}. This work was based on the application of semantic web techniques such as ontologies and rules for authentication and authorisation purposes in IoT. Specifically, IoT devices were used to gather information about the environment status. This information was used to perform the user's modelling and let him/her utilise certain services. Authors implemented the architecture and evaluated the resource consumption, scalability and authentication process confidence. Results reached an authentication confidence average of 78\%. However, as authors claimed, the system performance is highly related to the deployment context. Then, vast training is necessary to accurately model the users’ behaviour.

In conclusion, the existing continuous authentication proposals considering single- and multi-device scenarios do not combine the behaviour of users with different devices to infer additional behavioural patterns, as our work does. \tablename~\ref{tab:discussion} provides a global overview of the key aspects considered by each solution. As it can be appreciated, previous continuous authentication solutions obtained good performance during the authentication process. However, the ones achieving better performance are only focused on single-device scenarios. Therefore, our proposal goes beyond the state-of-the-art and improves some of the limitation found in current solutions. 

\section{Motivating example: Multi-device profiling}
\label{sec:motivating}

This section seeks to provide an illustrative example of the deficiencies of single-device continuous authentication as well as to motivate how multi-device systems could reduce or even mitigate these deficiencies.

Let's suppose a scenario based on a Smart Office where two employees, Bob and Alice, utilize their computers and smartphones to perform their daily tasks. Each device hosts a continuous authentication application that monitors and permanently authenticates the owner according to his/her behaviour \change{with applications, sensors, mouse, or keyboard, among others.}{. This behaviour is determined by the usage of the screen, sensors, and applications in the mobile device. Regarding PCs, other dimensions such as the keyboard, mouse, resource usage and application monitoring can be considered.}

In such Smart Office, Alice has been observing Bob when he used his smartphone, and she learned how to replicate his behaviour to impersonate Bob. \addtxt{This impersonation is achieved by repeating characteristics of Bob's activities, such as the way he holds the device, the speed at which he interacts with the screen and writes, or applications used and the application change patterns.} One day, while Bob is working on his computer, Alice takes advantage of Bob's distraction and gets his smartphone to access and obtain sensitive information. Since Alice replicated Bob's behaviour, the single-device continuous authentication application cannot detect that Alice is using the device, instead of Bob. In this scenario, only a multi-device continuous authentication system could detect that Bob's smartphone and computer \change{are used at the same time in different locations, which is physically impossible.}{are used exactly at the same time, something that is not typical in Bob's behaviour while working.}

Now, let's suppose that Bob installed and started using two new applications on his smartphone. It implies a smooth change in his behaviour, affecting the authentication accuracy of the single-device continuous authentication application. After that, the authentication application sometimes fails and does not detect Bob's behaviour as the owner (which means a high rate of false positives). This annoying situation can be reduced or mitigated by having a multi-device continuous authentication system that combines the behaviours of Bob with their devices in different time windows.

Finally, a multi-device continuous authentication system can also detect suspicious behaviours in one device compared to others' behaviour. As an example, multi-device could identify that recreational apps are used in the smartphone while the computer is being used for work. These situations are not detectable when single-device behaviour profiles are considered.

In conclusion, single-device continuous authentication solutions are capable of recognizing the device owner based on its behaviour. However, these solutions are sensitive to false-positive rates when small behaviour changes happen and can suffer impersonation attacks if the attacker knows the device owner's normal behaviour. \change{Having information from several devices improves the robustness and flexibility of the authentication system since it would be necessary to have access to several devices and mimic the user's behaviour in each of them.}{These drawbacks can be improved by a multi-device authentication system able to combine information from several devices. This combination would improve the authentication robustness and flexibility, since an attacker would need to mimic the user in all the devices to attack the system integrity.}

\section{AuthCODE architectural design} 
\label{sec:architecture}

\begin{figure}
	\centering
	\includegraphics[width=1\columnwidth]{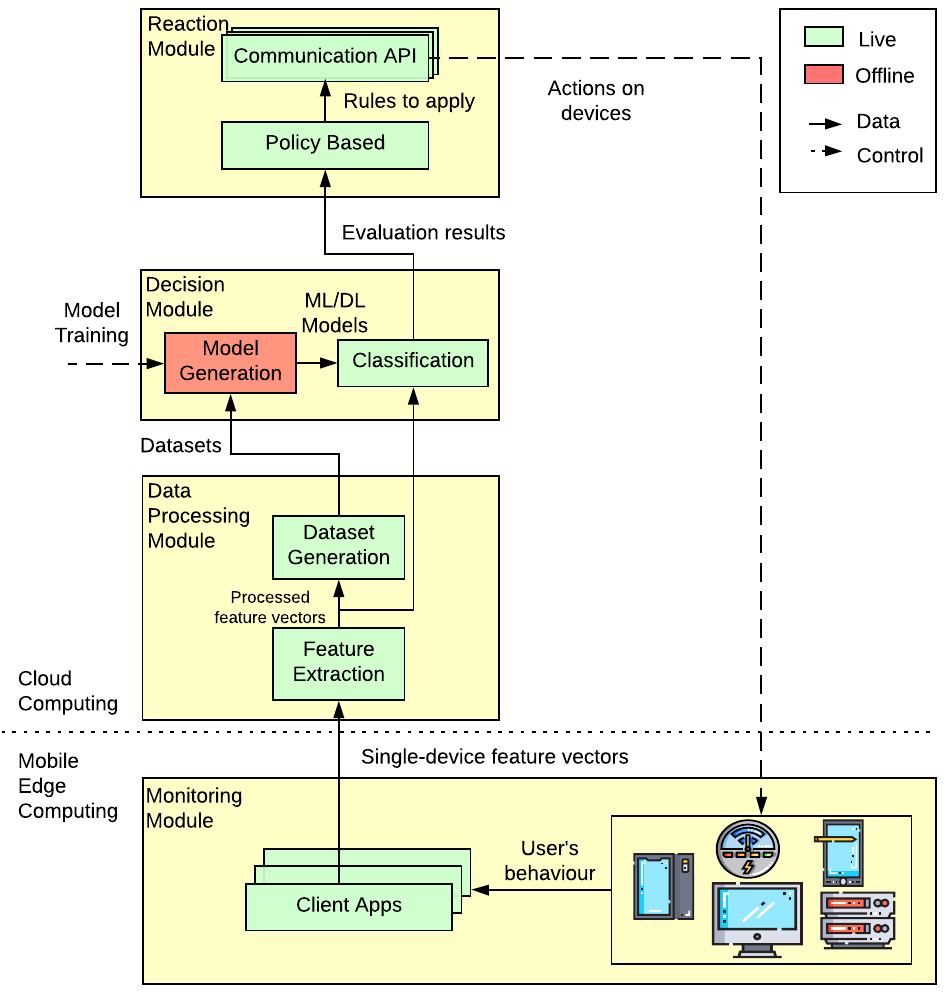}
	\caption{Design of the modules and components making up the AuthCODE architecture.} 
    \label{fig:design}
\end{figure}

This section describes the design details of our AI-based, privacy-preserving and multi-device continuous authentication architecture called AuthCODE. The architecture has been designed following a modular approach to allow flexible and dynamic modifications of its modules. In addition, AuthCODE combines the MEC, where users' sensitive data is managed and features are calculated to protect users' privacy, with the cloud computing, where computationally complex data processing and AI-base techniques are executed. This is a key characteristic, since sensitive data is maintained in the users' devices, and the performance of resource-constrained devices is not affected because AI-based models are trained and evaluated in the cloud. \figurename~\ref{fig:design} shows modules and components making up the AuthCODE architecture.

The AuthCODE architecture is composed of the following four modules:

\begin{itemize}
    \item \textit{Reaction}. Provides users with a final authentication score according to the outputs of the Decision module. Furthermore, it exposes interfaces with heterogeneous devices that enable global continuous authentication mechanisms for multi-device scenarios.

    \item \textit{Decision}. Hosts and executes AI-based techniques able to train and evaluate different models that will authenticate users based on their behaviour with multiple devices.

    \item \textit{Data Processing}. Filters, aggregates, and processes individual features acquired by the Monitoring module to generate relevant combined feature vectors making up the single- and multi-device behavioural datasets.

    \item \textit{Monitoring}. Monitors the data generated by users interacting with their heterogeneous devices and calculates single-device vectors of features that do not contain sensitive data. Once the feature vectors are calculated they are sent to the Data Processing module for further processing.
\end{itemize}

From bottom to up, \textit{Monitoring} is the lowest module of AuthCODE and it is composed of several applications (Client Apps). Each client app runs on top of a device and monitors the data generated due to the user's actions. To ensure the privacy of users' sensitive data, each app processes and aggregates sensitive data in different windows of time established by the administrator\addtxt{, preventing the user's actions from being reconstructed or sensitive information from being inferred}. After the aggregation, the app generates single-device vectors of features that do not contain sensitive data and send them periodically to the Data Processing module, which runs in the cloud computing, as explained in Section \ref{sec:deployment}. 

The \textit{Data Processing} module periodically receives single-device feature vectors from each app. The Feature extraction component filters, aggregates and processes single-device feature vectors to calculate processed relevant features (which could belong to one or more devices depending on the scenario) that model the user's behaviour. The aggregation process of this module is also performed periodically and considering different time windows, which are established by the administrator as well. Finally, the Data Processing module generates datasets modelling the user's behaviour with different devices. These datasets can contain single-device or multi-device features, depending on the scenario needs.

The \textit{Decision} module has two main tasks: 

\begin{itemize}
    \item Off-line training. The datasets generated by the previous module are used to train a set of models by using different ML and DL algorithms. The scenario requirement will decide if it is needed one model per device, one multi-device model, or both. The training process is performed by the \textit{Model Generation} component only once and during the system bootstrapping.
    
    \item Real-time evaluations. Periodically and once the models have been trained, the \textit{Classification} component evaluates the real-time features vectors against the models to provide an authentication score per model.
\end{itemize}

To conclude, the \textit{Reaction} module aggregates the different authentication scores and calculates a global one as well as provides interfaces with the devices enabling a global and non-invasive multi-device continuous authentication. For that, the \textit{Policy-based} component considers rules that establish the user' authentication level, This level decides proper security actions such as unlock a particular device without requiring additional credentials, lock the device, or ask for authentication credentials. Strict rules can be applied over devices managing sensitive data or performing critical tasks, while more permissive rules can be applied over devices with secondary roles. Finally, the \textit{Communication API} sends the previous security reactions to the different devices as well as configures the time windows sent to the Monitoring module to aggregate sensitive data and calculate features.

\section{AuthCODE deployment \& Datasets} 
\label{sec:deployment}

This section shows the implementation details of the AuthCODE architecture as well as the generation of our five datasets in a realistic multi-device scenario such as a Smart Office. In our Smart Office, five users interacted with their smartphones, tablets, laptops and desktop computers for 60 days. Below we provide the implementation details of the modules making up our architecture.

\subsection{Mobile Edge Computing}

The Monitoring module and its Client Apps have been deployed close to the end users, in the MEC. They are hosted by Smart Office devices or by third-parties, in case of resource constrained devices. This decision ensures the performance and privacy-preserving capabilities of AuthCODE.

\subsubsection{Personal computer devices}

We have implemented a client app for Windows, the most used desktop OS, and another app for Linux distributions based on Debian. The apps monitor the folowing dimensions: 1) mouse movements/events, 2) keyboard events and 3) applications/resources usage statistics. \tablename~\ref{tab:dataPC} shows the selected dimensions and the data acquired from each dimension. We used Python, specifically, the pyinstaller tool to generate both Windows and Linux executable, the pynput library to monitor the mouse and keyboard events, and psutil and pywin32 (only in Windows) to monitor the resources consumption and applications usage~\citep{pyinstaller,pypi}. Finally, a time window is used to aggregate data and calculate features, which are sent to the Data Processing module using a REST API. In our implementation, the time window is set to 60 seconds. The feature selection process is explained and justified in Section \ref{sec:experiments}.

\begin{table}
	\caption{Privacy-preserving features obtained from personal computers.}
	\centering
	\begin{tabular}{cl}
        \hline
			\textbf{Dimension} & \makecell[c]{\textbf{Features}} \\
		\hline
		\makecell[c]{Time\\(1 feature)} &\makecell[l]{- Vector timestamp.} \\
		\hline			
   		\makecell[c]{Keyboard\\(24\,002 features)} & \makecell[l]{- Keystroke \& word counter.\\ - Erasing keys percentage.\\ - Pressed keys histogram \\- Average \& standard deviation of \\time that keys are pressed/released.\\- Average \& standard deviation of  \\consecutive keystroke intervals .\\ - Number of written words \& length\\ histogram. \\- Digraphs typed (two consecutive \\keys) \& mean time to type the \\digraph.} \\
         \hline			
        \makecell[c]{Mouse\\(45 features)} & \makecell[l]{- Clicking speed average \& standard \\deviation per mouse button and left\\ double clicking.\\- Average mouse speed per direction.\\- Movement length histogram.} \\
        \hline		
        \makecell{Application\\and\\resource\\usage\\(17 features)} & \makecell[l]{- ID of the last and penultimate \\application used.\\- Active application counter average. \\- Counter of application changes.\\- CPU/RAM usage average and \\standard deviation.\\- Bytes transmitted \& received \\through the network\\ interfaces.} \\
        \hline
    \end{tabular}
	\label{tab:dataPC}
\end{table}

\subsubsection{Mobile devices}

We have also deployed a client app for smartphones and tablets running from Android 5.0 OS (API 21). We chose Android since it is the most used operating system on mobile devices. In this case, the monitored dimensions are 1) the application usage statistics, and 2) the device sensors. \tablename~\ref{tab:finalfeaturesandroid} 
shows the data extracted from the previous dimensions. We have used Android.app.usage class~\citep{AndroidDevelopers} to gather the application usage statistics and the Android.hardware.SensorEventListener interface to obtain the gyroscope and acceleromenter sensors data. To maintain a low resource consumption, we implemented a short-time service, which is triggered cyclically through the Android.app.AlarmManager. As in the personal computers case, the previous data is aggregated in time windows, and features are calculated and send to the Data Processing module through an REST API. In our implementation, the time window is set to 60 seconds. The feature selection process is explained in Section \ref{sec:experiments}.

\begin{table}
		\caption{Privacy-preserving features obtained from mobile devices.}
	\centering
	\begin{tabular}{cl}
		\hline	
		\textbf{Dimension} & \textbf{Features} \\
		\hline
		\makecell[c]{Time\\(1 feature)} &\makecell[l]{- Vector timestamp.} \\
		\hline
       \makecell[c]{Application \\ usage \\statistics\\(13 features)} & \makecell[l]{- Foreground application counters \\(number of different and total apps)\\ for the last minute and day. \\
         - Most common app ID and number of\\ usages in the last minute and day. \\
         - ID of the currently active app \\
         - ID of the last active app prior the \\current one.\\
         - ID of the application most frequently \\utilised prior to the current application.\\
         - Bytes transmitted \& received through\\ the network interfaces.} \\
		\hline
		\makecell[c]{Sensors \\ (Gyroscope \\and\\ Accelerometer)\\(40 features)} &\makecell[l]{- Average, maximum, minimum, varian-\\ce and peak-to-peak (max-min) of\\ X, Y, Z coordinates. \\
   		 \textit{- Magnitude = \(\sqrt[]{X^2+Y^2+Z^2}\)}} \\
		\hline	
	\end{tabular}
	\label{tab:finalfeaturesandroid}
\end{table}

\subsection{Cloud computing platform}

\begin{figure}
	\centering
	\includegraphics[width=\columnwidth]{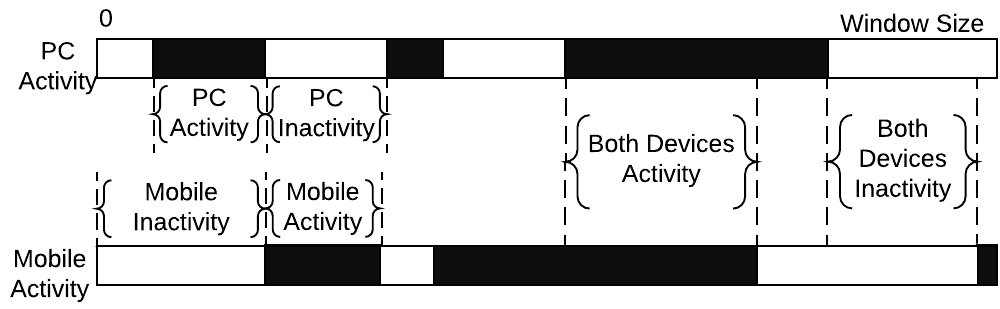}
	\caption{Explanatory diagram on the derived features from both user devices.}
    \label{fig:combined_scheme}
\end{figure}

    Due to storage and processing requirements, a private cloud platform hosts the Data Processing, Decision, and Reaction modules of AuthCODE. The Data Processing module exposes a REST API to receive single-device feature vectors from the Client apps. Periodically, AuthCODE follows the next steps to create our five datasets: 1) it processes and aggregates single-device feature vectors from the same user to generate multi-device feature vectors for every user using the user ID as label, 2) translates the features domains to make them suitable for ML/DL classifiers. \tablename~\ref{tab:combined_features} shows some of the most relevant multi-device features obtained after aggregating each device vector timestamps in a given time window. \figurename~\ref{fig:combined_scheme} illustrates the four different activity combinations that arise when two devices are involved: none of the devices is active, only one of them is active, or both devices are active simultaneously.  
These features are tested and validated in Section~\ref{sec:experiments}. After following the previous steps AuthCODE generated the following five datasets~\citep{datasets}:

\begin{itemize}
    \item \textit{Dataset 1}. Single-device behaviour profile obtained from the personal computer, comprising aggregated data about keyboard and mouse activity, as well as application usage statistics (see \tablename~\ref{tab:dataPC}).
    \item \textit{Dataset 2}. Single-device behaviour profile obtained from the mobile device, with application usage statistics (see \tablename~\ref{tab:finalfeaturesandroid}).
    \item \textit{Dataset 3}. Single-device behaviour profile with features computed from the sensors of the mobile device (see \tablename~\ref{tab:finalfeaturesandroid}).
    \item \textit{Dataset 4}. Multi-device behaviour profile combining the most relevant features of the mobile device and personal computer.
    \item \textit{Dataset 5}. Multi-device behaviour profile generated from the active/inactive intervals of both devices (see \tablename~\ref{tab:combined_features}).
\end{itemize}

\begin{table}
		\caption{Privacy-preserving multi-device features.}
	\centering
    
	\scalebox{.95}[0.95]{\begin{tabular}{c}
	    \hline
	    \textbf{Features (32 in total)}\\
		\hline
		Hour when the time window starts. \\
		\hline
		Weekday when the time window starts. \\
		\hline			
        Total number of vectors from PC. \\
        \hline
        Total number of vectors from mobile devices. \\
        \hline
        Number of changes pc-mobile. \\
        \hline
        Number of changes mobile-pc. \\
        \hline
        Number of minutes with activity of both devices. \\
		\hline
	    \makecell[c]{Mean, stdev, max, min of the \\PC Activity periods duration}.\\
	    \hline
	    \makecell[c]{Mean, stdev, max, min of the \\Mobile Activity periods duration}.\\
	    \hline 
	    \makecell[c]{Mean, stdev, max, min of the \\Both Devices Activity periods duration}.\\
	    \hline
	    \makecell[c]{Mean, stdev, max, min of the\\ PC Inactivity periods duration}.\\
	    \hline
	    \makecell[c]{Mean, stdev, max, min of the \\Mobile Inactivity periods duration}.\\
	    \hline
	    \makecell[c]{Mean, stdev, max, min of the \\Both Devices Inactivity periods duration}.\\
	    \hline
	\end{tabular}}
	\label{tab:combined_features}
\end{table}

Once the datasets are generated, the Decision module uses ML and DL algorithms to classify the users. Thus we had to implement a selected set of ML/DL models both for single- and multi-device profiles. To this end we used three well-known Python libraries: Scikit-learn, Keras and Pandas. Scikit-learn provides a wide variety of ML algorithms for both classification and anonaly detection. The Keras framework is widely used to implement and train DL models. And finally, the Pandas library was employed to manipulate and process data. Finally, the Reaction module deploys the Policy-based component with management rules, which are implemented in Python. Moreover, the Communication API component implements a REST API to send the user's authentication result and the actions to be performed to the Smart Office devices.

\section{Experiments}
\label{sec:experiments}

We measured the AuthCODE performance in terms of authentication accuracy and resource consumption. To accomplish this objective, the interactions of five individuals with their computers and mobile devices were collected for 60 days by means of the client apps detailed in Section~\ref{sec:deployment}.

\subsection{Comparing single- and multi-device authentication.}
\label{subsec:experimentA}

This experiment analysed and compared the classification accuracy of AuthCODE considering single- and multiple-device behaviour profiles. Additionally, it justified the list of features selected for both single- and multi-device profiles (see \tablename~\ref{tab:dataPC}, \ref{tab:finalfeaturesandroid} and \ref{tab:combined_features}). To analyse the previous aspects we considered Dataset 1-4 of Section~\ref{sec:deployment}.

Given the huge number of features of some of the datasets (+24\,000 features in Dataset 1) it was necessary a preliminary stage of feature selection. Our first step was to preprocess the three datasets discarding all the constant features and encoding each categorical one by using one-hot representation. Next, we chose RF and XGBoost~\citep{XGBoost} to perform an initial classification process of each single-device dataset for additional feature selection purposes. Both algorithms provide an estimation of the discriminative power of each feature. Additionally, they were chosen due to their ability to manage high number of features and their good performance dealing with imbalanced classes (some users have more activity than others). Each dataset was partitioned in 10-minute segments and 10\% segments were randomly chosen to create the test set. The previous and next segments for each selected segment were discarded to prevent data leakage in the training set due to correlation issues. Both algorithms were trained using each of the first three datasets, and the estimated discriminative of each feature was used to select a subset of features that comprised 95\% of the total importance. Finally, Dataset 4 was created by combining these three resulting datasets to include multi-device information. The four final datasets were then used to train a set of candidate ML algorithms besides RF and XGBoost: Naive Bayes, k Nearest Neighbours (k-NN), Support Vector Machine (SVM) and Multi-Layer Perceptron (MLP). Each training was carried out using a validation set randomly selected similarly as the test set. The corresponding validation set of each dataset was used to perform proper optimization of hyperparameters for each ML algorithm. The list of hyperparameters per ML algorithm is detailed in \tablename~\ref{tab:clasif_alg_hyp}. The performance metrics used to evaluate the models were the following (FPR: False Positive Rate, FRR: False Rejection Rate) :

\begin{equation}
    Precision=\frac{TP}{TP+FP}
\end{equation}
\begin{equation}
    Recall=\frac{TP}{TP+FN}
\end{equation}
\begin{equation}
    F1-Score=\frac{2 \times precision \times recall}{precision + recall}
\end{equation}
\begin{equation}
    FPR =\frac{FP}{FP+TN}
\end{equation}

\begin{table}
	\caption{Classification algorithms and hyperparameters tested.}
	\centering
    \begin{tabular}{m{1.5cm}m{6cm}}
        \hline
        \textbf{Algorithm} & \textbf{Hyperparameters} \\
        \hline
        Naive Bayes & No hyperparameter tunning required \\
        \hline
         k-NN & $k\in [3,20]$ \\
         \hline
         SVM & \makecell[l]{$C\in [0.01,100], gamma\in [0.001,10]$\\$kernel\in \{'rbf', 'linear', 'sigmoid','poly'\}$}\\
         \hline
         XGBoost & \makecell[l]{$lr \in [0.01,0.30], max\_depth \in [3,15]$\\ $min\_child\_weight\in [1,7], gamma \in [0,0.5]$\\$colsample\_bytree \in [0.3,0.7]$}\\
         \hline
         MLP & $layers\in [1,5], neurons\_layer\in [50,1000]$ \\
         \hline
         Random Forest & $Number\_of\_trees \in [50,1000]$\\
         \hline
    \end{tabular}
    \label{tab:clasif_alg_hyp}
\end{table}{}

\subsubsection{Single-device classification on personal computer data}

The preliminary preprocessing of Dataset 1 reduced its number of features from +24\,000 to 12\,160. Additionally, the timestamp was replaced by just the time of day. With this transformed dataset, XGBoost reached a slightly better performance than RF. Therefore, we used the features importance provided by XGBoost to selected the 150 most relevant ones, training with them every candidate classification model. This decision drastically reduced the dimensionality (from 12\,160 to 150 features) for future testing without loss of classification performance.

\begin{figure}
	\centering
	\includegraphics[trim=0cm 1cm 0cm 0cm, clip=true, width=\columnwidth]{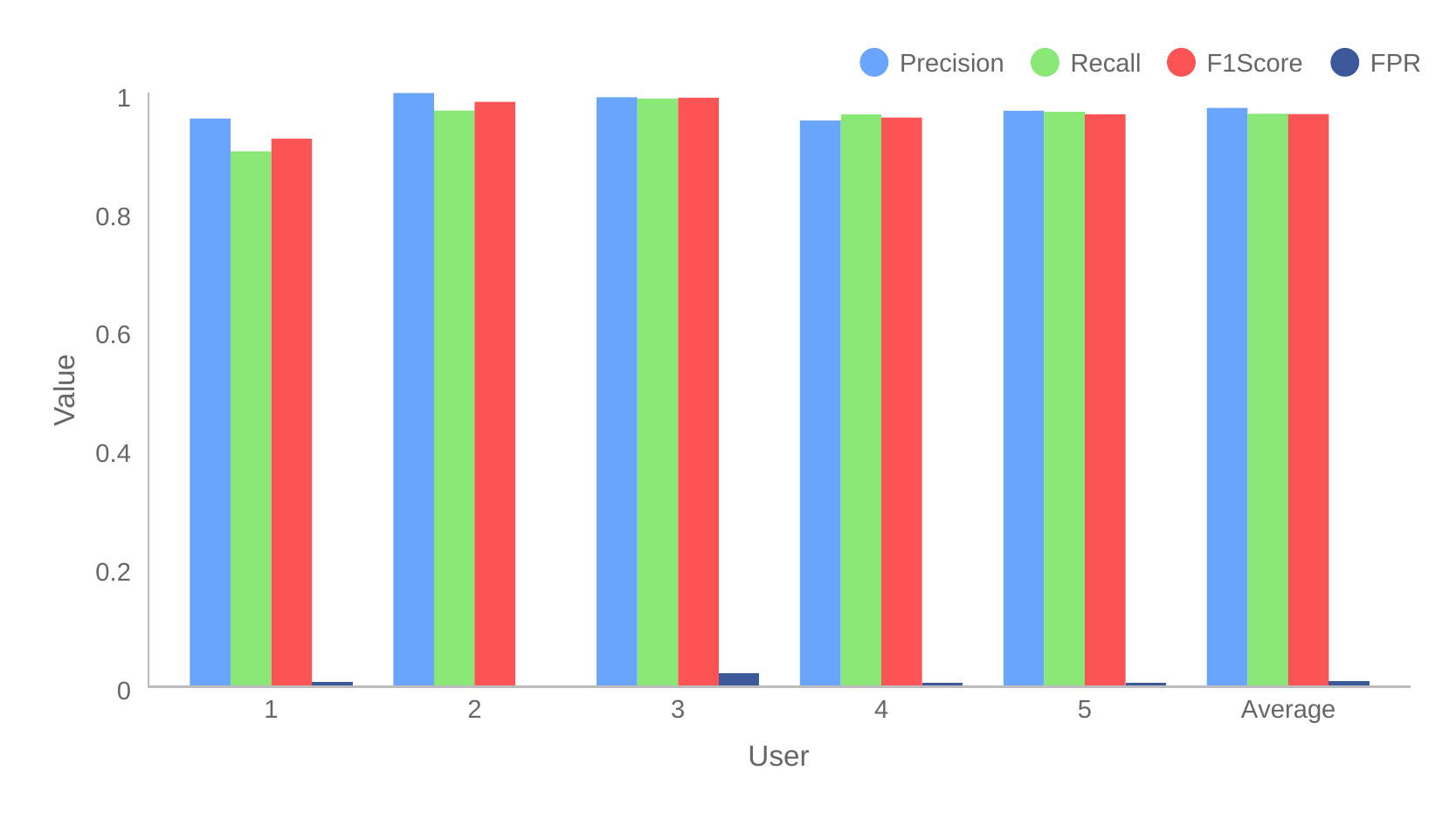}
	\caption{Classification performance achieved by MLP on Dataset 1 (PC data).}
    \label{fig:resultsPC}
\end{figure}

As it can be seen in \tablename~\ref{tab:clasif_expA}, MLP with one single hidden layer of 500 neurons obtained the best average classification performance, reaching a f1-score of 97.39\%. Furthermore, \figurename~\ref{fig:resultsPC} shows the results of identifying each of the five users from their personal computer usage. Additionally, MLP reached for every single user a precision, recall, and f1-score higher than 95\%, 90\% and 95\%, respectively. Regarding FPR, its values were smaller than 3\% for every user.

\subsubsection{Single-device classification on mobile device data}

We evaluated the classification performance of the RF and XGBoost models on application statistics data (Dataset 2) and sensor data (Dataset 3) separately. RF achieved better classification performance than XGBoost on both datasets. Based on the discriminating importance of each feature provided by RF, the 50 most important features of Dataset 2 and the 40 most important features of Dataset 3 were selected, reducing the number of features from 818 to 50, and from 88 to 40, respectively. 

\begin{figure*}
	\centering
	\begin{subfigure}{0.49\textwidth}
        \includegraphics[trim=0cm 1cm 0cm 0cm, clip=true,width=\textwidth]{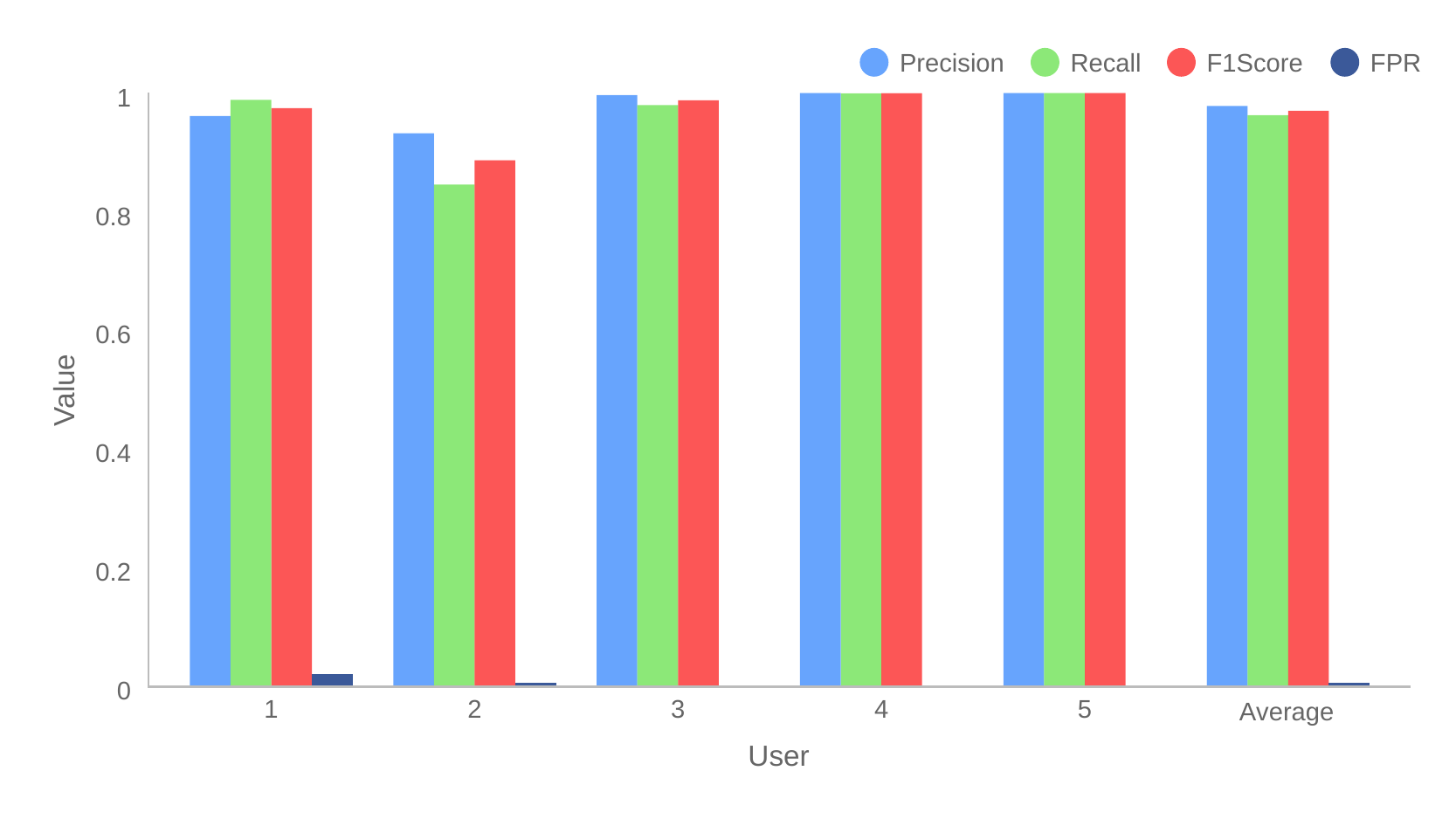}
        \caption{Classification performance achieved by RF on Dataset 2 (Application usage statistics).}
        \label{fig:resultsMobile_apps} 
	\end{subfigure}
	\begin{subfigure}{0.49\textwidth}
        \includegraphics[trim=0cm 1cm 0cm 0cm, clip=true,width=\textwidth]{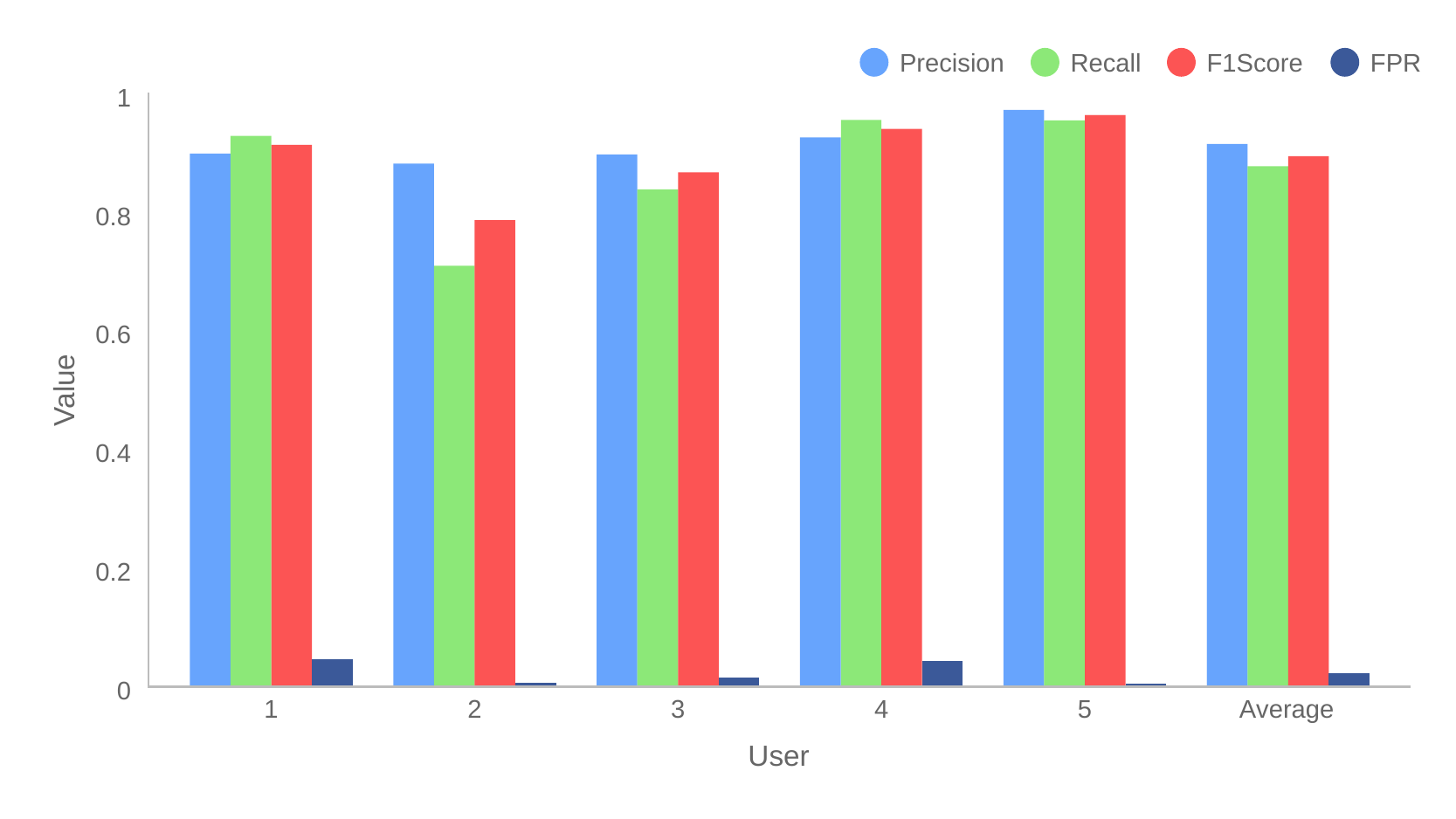}
        \caption{Classification performance achieved by XGBoost on Dataset 3 (Sensors data).}
        \label{fig:resultsMobile_sensors} 
	\end{subfigure}
	\caption{Users' behaviour classification in mobile devices.}
    \label{fig:resultsMobile}
\end{figure*}

Once selected the most discriminating features, \tablename~\ref{tab:clasif_expA} shows that RF, with \textit{Number\_of\_trees:} 500, obtained the best performance with Dataset 2, and XGBoost, with \textit{lr:} 0.25, \textit{max\_deph:} 10, \textit{min\_child\_weight:} 3, \textit{gamma:} 0.5 and \textit{colsample\_bytree:} 0.5, was the best model for Dataset 3. Additionally, \figurename~\ref{fig:resultsMobile} shows the classification results for each user by considering the previous models for each dataset.

\subsubsection{Multi-device classification}
\label{multi-device_classif}
\begin{figure}
	\centering
	\includegraphics[width=0.95\columnwidth]{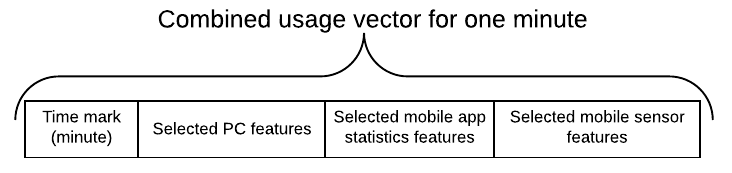}
	\caption{Structure of multi-device features vector making up Dataset 4.}
    \label{fig:combined_vector}
\end{figure}

We used Dataset 4 to evaluate whether the combination of the most discriminating features of personal computers (150) and mobile devices (50 for apps statistics and 40 for sensors) could improve the authentication results obtained in the previous two experiments. With that goal in mind, the previous features were grouped in time windows of one minute. In this way, we created feature vectors representing the activities of each user interacting with the two devices in the same minute. \figurename~\ref{fig:combined_vector} depicts a scheme of the morphology of the vector generated. If any of the devices (PC or mobile) has no activity in that minute, their features are established to 0. Only vectors with activity are generated.

The relatively low number of features (240) and the better class balance due to the combination of the other three datasets, made the multi-device Dataset 4 suitable to be used without changes to train the set of ML algorithms, including their hyperparameter optimization. \tablename~\ref{tab:clasif_expA} shows the best classification results for each algorithm on Dataset 4. The winner was XGBoost, with the following hyperparameters: \textit{lr}: 0.25, \textit{max\_depth:} 10, \textit{min\_child\_weight:} 5, \textit{gamma}: 0.1 and \textit{colsample\_bytree}: 0.7. 


The results of the previous three subsections demonstrated how single-device classification results can be improved by combining single-device features to create multi-device profiles. \tablename~\ref{tab:clasif_expA} compares the results of each one of the experiments of this section, demonstrating that the best results were obtained with multi-device classification and XGBoost with the following hyperparameters: \textit{lr}: 0.25, \textit{max\_depth:} 10, \textit{min\_child\_weight:} 5, \textit{gamma}: 0.1 and \textit{colsample\_bytree}: 0.7. 

\begin{table}
    \scriptsize
	\caption{Comparison of classification algorithms for single- and multiple-device behaviour profiles.}
	\centering
	\begin{tabular}{m{0.9cm}m{0.7cm}m{0.7cm}m{0.7cm}m{0.7cm}m{0.7cm}m{0.7cm}}
        \hline
    	\makecell[c]{\textbf{Model}} & \makecell[c]{Naive\\Bayes} & K-NN & \makecell[c]{SVM} & \makecell[c]{XG\\Boost} & \makecell[c]{MLP} & \makecell[c]{RF}\\
    	\hline
        \hline
        \multicolumn{7}{ |l| }{\cellcolor{lightgray}Dataset 1: Personal computer} \\ \hline
    	\textbf{Avg.} \textbf{Precis.} & 0.6977 & 0.9280 & 0.9593 & 0.9708 & \textbf{0.9752} & 0.9710 \\
         \hline
         \textbf{Avg.} \textbf{Recall} & 0.6425 & 0.9187 & 0.9532 & 0.9525 & \textbf{0.9727} & 0.9279 \\
         \hline
         \textbf{Avg.} \textbf{F1-Scr.} & 0.6054 & 0.9221 & 0.9561 & 0.9610 & \textbf{0.9739} & 0.9459 \\
         \hline
         \textbf{Avg.} \textbf{FPR} & 0.0951 & 0.0125 & 0.0092 & 0.0096 & \textbf{0.0072} & 0.0085\\
         \hline \hline
		\multicolumn{7}{ |l| }{\cellcolor{lightgray}Dataset 2: Applications usage statistics} \\ \hline
		\textbf{Avg.} \textbf{Precis.} & 0.8281 & 0.9473 & 0.9318 & 0.9466 & 0.9505 & \textbf{0.9775} \\
         \hline
         \textbf{Avg.} \textbf{Recall} & 0.8005 & 0.9368 & 0.9057 & 0.9390 & 0.9487 & \textbf{0.9534} \\
         \hline
         \textbf{Avg.} \textbf{F1-Scr.} & 0.7286 & 0.9418 & 0.9174 & 0.9421 & 0.9504 & \textbf{0.9670} \\
         \hline
         \textbf{Avg.} \textbf{FPR} & 0.0523 & 0.0082 & 0.0112 & 0.0118 & 0.0085 & \textbf{0.0057}\\
         \hline \hline
         \multicolumn{7}{ |l| }{\cellcolor{lightgray}Dataset 3: Sensors} \\ \hline
         \textbf{Avg.} \textbf{Precis.} & 0.2915 & 0.7701 & 0.8682 & \textbf{0.9242} & 0.8439 & 0.9113 \\
         \hline
         \textbf{Avg.} \textbf{Recall} & 0.2751 & 0.7327 & 0.8171 & \textbf{0.8871} & 0.6502 & 0.8547 \\
         \hline
          \textbf{Avg.} \textbf{F1-Scr.} & 0.2360 & 0.7483 & 0.8385 & \textbf{0.9036} & 0.7328 & 0.8783 \\
         \hline
         \textbf{Avg.} \textbf{FPR} & 0.1933 &  0.0618 & 0.0355 & \textbf{0.0216} & 0.0882 & 0.0271\\
         \hline \hline
         \multicolumn{7}{ |l| }{\cellcolor{lightgray}Dataset 4: Muti-device} \\ 
         \hline
		 \textbf{Avg.} \textbf{Precis.} & 0.7425 & 0.9342 & 0.9606 & \textbf{0.9932} & 0.9712 & 0.9799 \\
         \hline
         \textbf{Avg.} \textbf{Recall} & 0.7577 & 0.9362 & 0.9671 & \textbf{0.9933} & 0.9476 & 0.9694 \\
         \hline
         \textbf{Avg.} \textbf{F1-Scr.} & 0.6862 & 0.9351 & 0.9637 & \textbf{0.9933} & 0.9591 & 0.9743 \\
         \hline
         \textbf{Avg.} \textbf{FPR} & 0.0544 & 0.0115 & 0.0062 & \textbf{0.0023} & 0.0064 & 0.0045\\
         \hline
    \end{tabular}
	\label{tab:clasif_expA}
\end{table}

Analysing the previous results, two main conclusions are obtained. On the one hand, AuthCODE classifies and authenticates users in single-device scenarios such as PC and mobile devices with 97.39\%, 96.70\% and 90.36\% average f1-score and 0.72\%, 0.57\% and 2.16\% average FPR for computer, mobile application statistics and mobile sensors, respectively. On the other hand, the proposed multi-device behaviour profile dataset improves the authentication results of single-device datasets, obtaining 99.33\% average f1-score and 0.23\% average FPR. It is interesting to note that the FPR with multi-device achieved a 69.33\%, 59,65\% and 89,35\% improvement for computer, mobile applications and mobile sensors respectively.

\subsection{Measuring the time impact in multi-device authentication.}

In the previous section, the user identification improvement obtained by using multi-device behaviour profiles collected in 1-minute time windows was evaluated. In this experiment we analysed the impact of using a sequence of such 1-minute snapshot vectors, seen as the evolution of a user's activity over the time. The temporal information carried in the sequence can be represented in different ways. We studied two approaches: authentication by means of derived temporal activity features and by means of time series of vectors.

\subsubsection{Derived features classification}

In this experiment we evaluated whether it was possible to identify users according to the time they spent interacting with their devices. For this purpose, we labelled each 1-minute vector from Datasets 1-3 with its associated device (computer or mobile). Subsequently, the resulting vectors were sorted by their timestamps and grouped in a variety of time windows (1 hour, 3 hours, 6 hours, 12 hours and 24 hours). For each time windows size, a dataset was created with features about the usage periods of the devices present in the window and the device changes made by the user (see \tablename~\ref{tab:combined_features}). Figure~\ref{fig:combined_vector} illustrates a window containing a group of vectors belonging to two devices. It can be noticed how the user switches between the two devices and the periods where there is activity on just one device, both devices or even no activity at all.

Our set of ML algorithms were trained with this dataset, including a hyperparameter optimization procedure. RF was the model that achieved the best performance (number of trees = 200). \tablename~\ref{tab:combined_classification} lists the results for each selected time windows. The classification results improve as window size increases, ranging from 72.59\% f1-score and 6.21\% FPR with 1-hour window to 92.45\% f1-score and 2.15\% FPR with 24-hour window. These results make sense, since the larger the window, the more interactions contains, providing more information to differentiate clearly each user's behaviour. 

Although the performance achieved did not improve the results obtained in Section~\ref{subsec:experimentA}, this experiment demonstrated the potential of using derived features to identify different users based on their device usage routines.

\begin{table}
	\caption{Derived usage features classification results using RF (Dataset 5).}
	\centering
	\begin{tabular}{cccccc}
        \hline
		\makecell[c]{\textbf{Time}\\ \textbf{window}} & \textbf{1 h} & \textbf{3 h} & \textbf{6 h} & \textbf{12 h} & \textbf{24 h}\\
		\hline
		\textbf{\makecell[c]{Average\\ Precision}} & 0.7298 & 0.7893 & 0.8005 & 0.8224 & 0.9310 \\
         \hline
         \textbf{\makecell[c]{Average\\ Recall}} & 0.7291 & 0.7883 & 0.8005 & 0.8175 & 0.9253 \\
         \hline
         \textbf{\makecell[c]{Average\\ F1-Score}} & 0.7259 & 0.7884 & 0.7980 & 0.8193 & 0.9245 \\
         \hline
         \textbf{\makecell[c]{Average\\ FPR}} & 0.0621 & 0.0498 & 0.0480 & 0.0491 & 0.0215 \\
         \hline
    \end{tabular}
	\label{tab:combined_classification}
\end{table}

\subsubsection{Time window processing using LSTM}

In this experiment we leveraged the ability of DL, and more specifically Long-Short Term Memory Neural Networks (LSTM), to learn complex patterns in sequences of vectors obtained from the activity data from the users' devices. LSTM is effective at capturing long-term temporal dependencies; therefore, our aim was to evaluate whether it was able to use the temporal information to improve the level of authentication reached in the experiments of Section\ref{subsec:experimentA}. In those experiments we created Dataset 4 by aggregating 1-minute time windows of multi-device activity data in vectors of 240 features, obtaining an average f1-score of 99.33\% and FPR of 0.23\% with XGBoost. The same Dataset 4 was subsequently processed to be used as input of a selection of LSTM architectures in a variety of configurations. The steps we carried out were the following:

\begin{itemize}

    \item Dataset 4 includes 60 days of monitoring data but a some of the first and last days do not contain data from all users. Therefore, we selected a continuous period of 40 days in which the five users had activity data.
    \item The feature vectors were sorted by the minute in which they were generated. To make explicit the lack of activity in the sequence, every minute without activity had -1 in the rest of the vector features.
     \item We used the numpy to work with array views and generate a dataset of sequences of a given size. The numpy function \textit{stride\_tricks.as\_strided} is particularly useful, allowing us to obtain a sliding window view of our dataset. With this function we obtained 10 separate views of the dataset for a range of sliding window sizes (2, 5, 10, 20, 30, 60, 90, 120, 240 and 360 minutes).
\end{itemize}

\begin{figure}
	\centering
        \includegraphics[width=0.8\columnwidth]{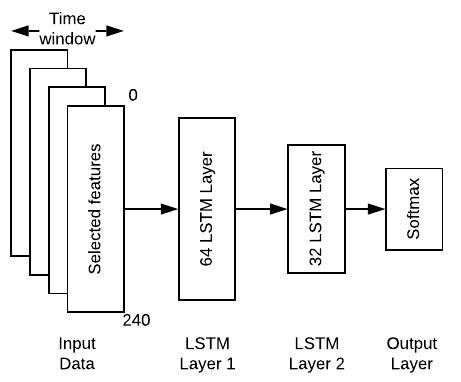}
        \caption{LSTM network architecture. }
        \label{fig:LSTM} 
\end{figure}

Our main architecture was composed of an optional 1D-convolutional layer (Conv1D), as suggested by some works in the literature~\citep{conv_LSTM1,conv_LSTM2}, followed by a number of LSTM layers (from 1 to 4) with different number of nodes (from 16 to 256) and a 5-node fully connected softmax output layer. All these configuration values together with the use of batch normalization and the dropout ratio in each layer, were hyperparameters to be optimized. Every dataset view was split into training/validation/test subsets. Each view was used as input during the hyperparameter optimization procedure. The validation and test datasets were composed of the sequences belonging to 7 days each: days 26 to 33 and 34 to 40 respectively.
The model configuration that achieved the higher performance was a 2-layer LSTM with 64 and 32 nodes, without Conv1D layer, no batch normalization and 20\% dropout. Figure~\ref{fig:LSTM} shows an scheme of this network.

\tablename~\ref{tab:resul_LSTM} lists the average precision, recall and f1-score when classifying behaviour data using different time window sizes. The table shows how as the time window size was increased, in general, the classification results were better. The lowest average f1-score and FPR (2 minute window) were 82.38\% and 3.81\% respectively, and they improved with window size until they reached 99\% and 0.37\% when the time window is 60 minutes. For greater windows the metrics stabilised in similar values, even when the window is six times larger (360 minutes).

\begin{table*}
	\caption{Temporal windows classification results using LSTM network (Dataset 6).}
	\centering
	\begin{tabular}{ccccccccccccc}
        \hline
		\makecell[c]{\textbf{Time Window (minutes)}} & \textbf{2} & \textbf{5} & \textbf{10} & \textbf{20} & \textbf{30} & \textbf{60} & \textbf{90} & \textbf{120} & \textbf{180} & \textbf{360} \\
		\hline
		\textbf{\makecell[c]{Average Precision}} & 0.9070 & 0.9532 & 0.9738	& 0.9794 & 0.9883 & 0.9971 & 0.9948 & 0.9984 & 0.9989 & 0.9988 \\
         \hline
         \textbf{\makecell[c]{Average Recall}} & 0.7584 & 0.8380 & 0.8453 & 0.9390 & 0.9538 & 0.9840 & 0.9777 &	0.9890 & 0.9932 & 0.9907 \\
         \hline
         \textbf{\makecell[c]{Average F1-Score}} & 0.8238 &	0.8898 & 0.9029 & 0.9579 &	0.9698 & 0.9902 & 0.9856 & 0.9934 & 0.9959 & 0.9945 \\
         \hline
         \textbf{\makecell[c]{Average FPR}} & 0.0381 & 0.0202 & 0.0199 & 0.0086 & 0.0065 & 0.0037 & 0.0038 & 0.0034 & 0.0028 & 0.0026 \\
         \hline
    \end{tabular}
	\label{tab:resul_LSTM}
\end{table*}

From these results some conclusions can be drawn. As it can be noticed in \tablename~\ref{tab:combined_classification} and \tablename~\ref{tab:resul_LSTM}, LSTM-based classification using Dataset 4 view as a dataset of sufficient large vector sequences, improved the results obtained when using Dataset 5 and RF. Also in \tablename~\ref{tab:resul_LSTM}, we can observe how performance improves as the time window increases. After the 60 minute window, the results stabilised around 99\% for f1-score and 0.30\% for FPR. So, it can be a good trade of between window size and performance. These results can be applied as a complement to the one-minute multi-device vector classification, giving an additional temporal information to the authentication process.

\subsection{Resource Consumption}

This experiment measured the resource consumption of client apps, running on smartphones and computers, as well as the ML/DL models of AuthCODE running on our server. For each testing device (see \tablename~\ref{tab:resource_consumption_devices}), relevant resources such as battery, memory, storage and processing, were monitored to check the impact of AuthCODE. Resources whose consumption is dynamic such as battery, memory, or CPU were monitored during 10 days, averaging the measurements of that period to calculate final results. Resources whose consumption is static, such as storage, were measured once the AuthCODE components were deployed.

\begin{table}
	\caption{Device specification of resource consumption tests.}
	\centering
	\begin{tabular}{cccc}
        \hline
		\textbf{Device} & \textbf{Processor} & \makecell[c]{\textbf{Mem.}\\\textbf{(GB)}}& \makecell[c]{\textbf{Battery}\\\textbf{(mAh)}}\\
		\hline
          \makecell[c]{Laptop:\\HP 15-bs00x} & \makecell[c]{Intel i7-7500U \\ 4 Cores @ 2.7 GHz} & 8 & 2850 \\
         \hline
           \makecell[c]{Laptop:\\Acer Nitro \\5 AN51} & \makecell[c]{Intel i7-7700HQ\\ 4 Cores @ 2.8 GHz} & 8 & 3270 \\
         \hline
           \makecell[c]{Smartphone:\\Xiaomi \\Pocophone} & \makecell[c]{Snapdragon 845\\ 8 Cores @ 2.8 GHz} & 6 & 4000 \\
         \hline
		 \makecell[c]{Smartphone:\\Huawei \\P10 Lite} & \makecell[c]{Kirin 658\\ 8 Cores @ 1.7 GHz} & 4 & 3000 \\
         \hline
          \makecell[c]{Cloud \\Server} & \makecell[c]{Intel Xeon E5-2697v4 \\ 18 Cores @ 2.3 GHz}  & 64 & - \\
         \hline
    \end{tabular}
    \vspace{1ex}
	\label{tab:resource_consumption_devices}
\end{table}

\subsubsection{Client Apps Consumption}

Battery, memory, storage and processing are the most critical resources of constrained resource devices such as laptops and smartphones. This experiment aimed to measure the impact of our client apps in the hosting devices resources.

\begin{itemize}
    \item \textit{Battery}. In average, the client app consumed 167 mAh ($\approx$4\%) and 201 mAh ($\approx$6\%) of the Xiaomi and Huawei battery devices, respectively. In terms of laptops, $\approx$4.75 mAh ($<$1\%) and $\approx$6.54 mAh ($<$1\%) of the HP and Acer batteries respectively were consumed.
    
    \item \textit{Memory}. In both laptops (HP and Acer), the client app consumed $\approx$25 MB. In contrast, for both mobile devices (Xiaomi and Huawei), the client app used $\approx$104 MB.

    \item \textit{Storage}. The client app executable file occupied 6.50 MB in the personal computers, and 15 MB in the smartphones (regardless the device model). The amount of raw data needed to compute a single-device feature vector is temporarily stored in the devices, having a size of $\approx$50 KB. Once the single-device feature vector is sent to the server, this storage is released.
    
    \item \textit{Processing}. In both laptops (HP and Acer), the average daily CPU usage ranged between 2\% and 5\%. In both smartphones (Xiaomi and Huawei), daily CPU usage remained under 1\%.

\end{itemize}

Based on the previous results, it can be concluded that neither computer client nor mobile client apps have a significant impact on the device resource consumption. Therefore, our client apps are suitable even for resource constrained devices.

\subsubsection{ML/DL Model Consumption} 

This experiment measured the resource and time consumption of the AuthCODE modules deployed in our server to train and evaluate our ML and DL models.

\begin{itemize}

    \item \textit{Time}. The average time needed to process each feature vector and evaluate it varied from 1.5 to 2 seconds. The measured process included the filtering and selection of features, the grouping of vectors in a 60-minute window, and the vector evaluation using XGBoost and the 60-minute LSTM network.

    \item \textit{Memory}. Once trained the models, they were loaded in memory and utilised to evaluate live users' vectors in real time. The memory usage of XGBoost and LSTM was 4.75 MB and 49.50 MB.
    
    \item \textit{Storage}. The Python scripts implementing Data Processing, Decision and Reaction modules of AuthCODE had a size of $\approx$6 KB. The generated models had a size of 2.05 MB for the XGBoost classifier (.pickle format) and 1.1MB for the LSTM models (.h5 format). In total, 3.15 MB of the server storage were used.
    
    \item \textit{Processing}. When the Data Processing module received features to evaluate the model and authenticate users, AuthCODE utilised $\approx$3\% in average.

\end{itemize}

In conclusion, we have proven that the AuthCODE modules makes efficient use of the processing, memory, storage, and battery resources of heterogeneous devices such as mobile devices, computers, and servers. In addition we have shown that the time required to evaluate and authenticate a given user remains in about 2 seconds, which is acceptable for our continuous authentication scenario prototype. Finally, it is important to keep in mind that the resource consumption optimization task is not the main objective of this work.

\section{Conclusions and Future Work} \label{conclusions}

This paper presents AuthCODE, a multi-device continuous authentication architecture, deployed in the MEC and cloud infrastructures, that utilises ML and DL techniques to authenticate users according to their behaviour. AuthCODE proposes a list of privacy-preserving and  multi-device features that combine the user's interactions with different devices. The relevance of the previous features as well as the improvement of multi-device profiles compared to single-ones have been validated in a Smart Office scenario, where we generated and published five behavioural datasets. Several experiments over the previous datasets demonstrated that multi-device profiles improved the authentication accuracy and FPR of solutions based on single-device profiles, reaching an f1-score of 99.33\% and a FPR of 0.23\% with XGBoost. Threrefore, our multi-device approach improves by 69.33\%, 59,65\% and 89,35\% the FPR obtained by computer, mobile applications and mobile sensors separately.

Additionally, the inclusion of temporal information in the form of vector sequences provides a further improvement in the authentication performance of the single-vector models, allowing the identification of complex behaviour patterns associated to each user. With this approach, an LSTM achieved an f1-score of 99.02\% and a FPR of 0.37\% with a 60-minute sequence of vectors. To conclude, several experiments also demonstrated the suitability of the proposed solution in terms of resource consumption at the mobile edge and cloud computing paradigms.

\addtxt{However, some limitations of the current proposal should be also commented. First, as user's behaviour is evaluated every minute, there is a short period of time in which an attacker could make use of the device without being detected. Besides, as ML/DL classification algorithms have been applied, user identification performance depends on how different each user's behaviour is from other users, and there can be model scalability problems if the number of users increases too much. Then, anomaly detection algorithms will be tested in the future. Finally, processing and evaluation is carried out in a centralized server, which can lead to scalability, availability or security issues. These limitations motivate to continue researching in multi-device continuous authentication solutions.}

As future work, we plan to evaluate the authentication accuracy of AuthCODE with more users and new experiments aiming to detect common behaviours by \addtxt{using new ML/DL algorithms and} filtering actions per type of application. Furthermore, we will extend the use case implementation by considering IoT devices, other operating systems, and new dimensions such as writing patterns or network traffic statistics. It will allow us to generate and release novel datasets, useful for researchers and scientist to keep improving the multi-device continuous authentication challenge.

\section*{Declaration of Competing Interest}
The authors declare that they have no known competing financial interests or personal relationships that could have appeared to influence the work reported in this paper. 

\section*{CRediT authorship contribution statement}

\textbf{Pedro Miguel S\'anchez S\'anchez.} Methodology, Writing - original draft, Data curation, Software. 
\textbf{Lorenzo Fern\'andez Maim\'o.} Methodology, Writing - Review \& Editing. Formal analysis.
\textbf{Alberto Huertas Celdr\'an.} Methodology, Conceptualization, Writing - Review \& Editing.
\textbf{Gregorio Mart\'inez P\'erez:} Supervision, Project administration, Funding acquisition. 

\section*{Acknowledgment}

This work has been partially supported by Armasuisse S+T with project CYD-C-2020003, by the University of Zürich UZH, and by the European Union Horizon 2020 Research and Innovation Program under grant agreement No. 830927, namely the H2020 Concordia Project. Special thanks to all those voluntaries who installed the client applications: Oscar Fern\'andez, Pedro A. S\'anchez, Francisco J. S\'anchez, Pantaleone Nespoli, Mattia Zago, Sergio L\'opez, Eduardo L\'opez, Manuel Gil, Jos\'e M. Jorquera, Javier Pastor and Gregorio Mart\'inez.

%
%
\bibliographystyle{cas-model2-names}
\bibliography{references}

\newpage

\bio{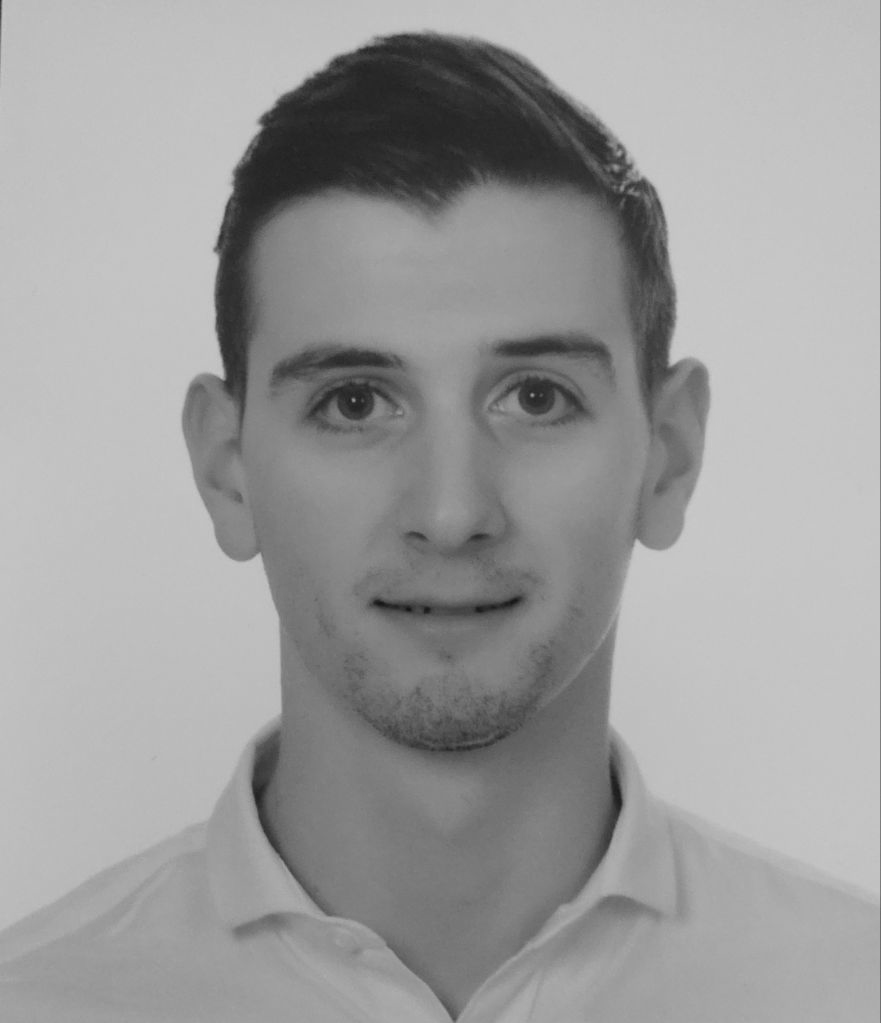}Pedro Miguel S\'anchez S\'anchez received the M.Sc. degree in computer science from the University of Murcia. He is currently pursuing his PhD in computer science at University of Murcia. His research interests are focused on continuous authentication, networks, 5G, cybersecurity and the application of machine learning and deep learning to the previous fields.\endbio

\bio{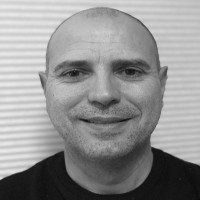}Lorenzo Fern\'andez Maim\'o received the M.Sc. and Ph.D. degrees in computer science from the University of Murcia. He is currently an Associate Professor with the Department of Computer Engineering, University of Murcia. His research interests primarily focus on machine learning and deep learning applied to cybersecurity, and computer vision.\endbio

\bio{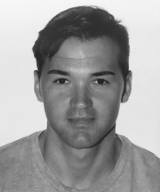}Alberto Huertas Celdr\'an received the M.Sc. and Ph.D. degrees in computer science from the University of Murcia, Spain. He is currently a postdoctoral fellow associated with the Communication Systems Group (CSG) at the University of Zurich UZH. His scientific interests include medical cyber-physical systems (MCPS), brain–computer interfaces (BCI), cybersecurity, data privacy, continuous authentication, semantic technology, context-aware systems, and computer networks.\endbio

\bio{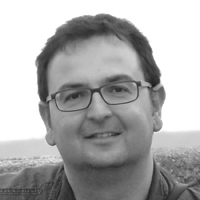}Gregorio Mart\'inez P\'erez is Full Professor in the Department of Information and Communications Engineering of the University of Murcia, Spain. His scientific activity is mainly devoted to cybersecurity and networking, also working on the design and autonomic monitoring of real-time and critical applications and systems. He is working on different national (14 in the last decade) and European IST research projects (11 in the last decade) related to these topics, being Principal Investigator in most of them. He has published 160+ papers in national and international conference proceedings, magazines and journals.\endbio

\end{document}